\documentclass[prb,reprint,notitlepage,amsmath,amsfonts,amssymb,floatfix]{revtex4-2}
\usepackage{graphicx}
\usepackage{bm}
\usepackage{color}
\usepackage{subcaption}
\usepackage{physics}
\usepackage{import}
\usepackage{gensymb}

\begin{document}   

\title{Combining complex conjugation, time-reversal, and spin-flip symmetry projection of coupled cluster wave functions}

\author{Ruiheng Song}
\email{rs84@rice.edu}
\affiliation{Department of Chemistry, Rice University, Houston, TX 77005-1892, USA}

\author{Thomas M. Henderson}
\affiliation{Department of Chemistry, Rice University, Houston, TX 77005-1892, USA}
\affiliation{Department of Physics and Astronomy, Rice University, Houston, TX 77005-1892, USA}

\author{Gustavo E. Scuseria}
\affiliation{Department of Chemistry, Rice University, Houston, TX 77005-1892, USA}
\affiliation{Department of Physics and Astronomy, Rice University, Houston, TX 77005-1892, USA}
\date{\today}

\begin{abstract}
Complex conjugation symmetry breaking and restoration generates two non-orthogonal configurations at the Hartree-Fock level that can capture static correlation naturally. In conjunction with broken spin-symmetry coupled cluster theory, the symmetry-projected wave function shows good agreement with full configuration interaction in beryllium hydride insertion, lithium fluoride dissociation, and symmetric stretching of tetrahedral H$_4$. By adding spin flip projection, we can also recover time reversal symmetry in the same coupled cluster framework.  We also show results including point group symmetry projection.
\end{abstract}

\maketitle

\section{Introduction}

The wave functions we use in quantum chemistry are typically real-valued.  That is, the molecular orbital coefficients in Hartree-Fock or density functional theory and the wave function amplitudes in methods such as coupled cluster theory are generally chosen to be real.  Choosing to work with real quantities simplifies implementation, since we do not need to keep track of complex conjugates, and also reduces storage requirements.  More precisely, these wave functions possess complex conjugation symmetry, which is essentially a realization of the fact that in a basis in which the Hamiltonian matrix is real, its eigenvectors can be chosen real.  We note that when we use a plane wave basis, the basis itself is not real and consequently neither are the orbital coefficients, but this does not imply that calculations in a plane wave basis necessarily break complex conjugation symmetry.

On the other hand, while complex wave functions are infrequently used in atomic orbital-based quantum chemistry, it has long been known that one can break complex conjugation symmetry at the mean-field level to obtain, in principle, variationally superior energies.\cite{Pople1977, Fukutome1981, Stuber2003} Indeed, research in the past several years has emphasized that a complex conjugation-broken wave function has a more flexible form in solving electronic structure problems.  For example, many complex determinants smoothly connect different solutions.\cite{MHG2015, Lena2018} Holographic Hartree-Fock provides continuous solutions across all geometries.\cite{ATom2018} Complex conjugation symmetry breaking is particularly important for general Hartree-Fock (GHF) wave functions which break both $S^2$ and $S_z$ symmetry since it is only on breaking complex conjugation symmetry that GHF permits a description of non-coplanar magnetism.\cite{Tom2018}

The fact remains, however, that when complex conjugation is a symmetry of the Hamiltonian, the wave function in a real basis should be real or, more precisely, should be an eigenstate of the complex conjugation operator $K$:
\begin{equation}
    K | \Psi \rangle = e^{\mathrm{i}\theta} | \Psi \rangle
\end{equation}
where the phase $\mathrm{e}^{\mathrm{i} \, \theta}$ can always be chosen to be $\pm 1$ by multiplying $|\Psi\rangle$ by the appropriate and physically inconsequential global phase factor.

Previous work has established many symmetry-projected wave function methods including Hartree-Fock,\cite{Hend1974, Carlos2012} configuration interaction,\cite{takashi2016} coupled cluster theory,\cite{Qiu2017, Qiu2018, Song2022, Takashi2018} and perturbation theory.\cite{takashi2019}  Most of this work has focused on the projection of spin symmetry, although the projection of number symmetry is frequently used in nuclear physics.\cite{Schmid2004}  Complex conjugation symmetry is a somewhat different beast.  It is a discrete symmetry, and because the complex conjugation operator is anti-unitary it does not have an observable quantum number. The implementation of complex conjugation projection with post-HF methods remains unexplored.

In this work, we review the complex conjugation operator and implement the complex conjugation projection of Hartree-Fock and coupled cluster theory. In addition, we add spin flip projection to recover time-reversal symmetry. We test these methods on small molecules with complex restricted Hartree-Fock, unrestricted Hartree-Fock (UHF), and generalized Hartree-Fock, as well as with coupled cluster with single and double excitations (CCSD) based on all of these mean-field references.  We aim to show that complex conjugation projection can be a useful tool for multi-reference problems.  We also show a few pitfalls which might arise, since frequently projecting one symmetry also projects or at least approximately projects others, and we show how we can circumvent these difficulties by adding extra projectors to cleanly disambiguate states.  In particular, projecting complex conjugation can also restore point group symmetry, and to have full control over the final states we can include point group symmetry projection along with all the others.

\section{Background}
Before we can discuss complex conjugation projection, we must briefly review the complex conjugation and time reversal operators.  Readers familiar with these ideas may skip the following subsections, though we use them to establish our notation.

\subsection{The Complex Conjugation Operator}
The complex conjugation operator is an anti-unitary operator which satisfies\cite{Wigner1960}
\begin{align}
\langle K \, \psi | \phi \rangle &= \langle \psi | K^+ \, \phi \rangle^*,
\label{Eqn:DefK}
\\
K^+ \, K &= K \, K = K \, K^+ = 1.
\end{align}
Here, we use $K^+$ for the adjoint of the complex conjugation operator to distinguish it from the Hermitian adjoint.  Note that the standard bra-ket notation is not particularly well equipped to handle antiunitary operators like $K$.  A symbol such as $\langle \phi|K|\psi\rangle$ is somewhat ambiguous, as one could understand it to mean acting $K$ to the right or the left.  Generally, these two operations will yield different results, so when there is any risk of ambiguity we will include $K$ inside the bra or ket, as we have done in Eqn. \ref{Eqn:DefK}.

When acting on a complex number, $K$ is just a simple complex conjugation. However, its action on a quantum state is slightly different.  This is perhaps easiest to see by way of an example.  Suppose that $|\Psi\rangle$ is a single determinant with real atomic orbitals and with complex molecular orbital coefficients $\boldsymbol{C}$.  Then $K |\Psi\rangle$ is also a single determinant, but its molecular orbital coefficients are $\boldsymbol{C}^\star$:
\begin{equation}
K |\Psi(\boldsymbol{C})\rangle = |\Psi(\boldsymbol{C}^\star)\rangle.
\end{equation}
We can always find a unitary transformation $R_K$ to do the same, so that we can write
\begin{equation}
K |\Psi \rangle = R_K |\Psi \rangle.
\end{equation}
We will use the operator in this form in our projected wave function methods.

\subsection{The Time Reversal and Spin Flip Operators}
We now turn to the spin flip operator $F=e^{i\pi S_y}$ and time reversal operator $\mathcal{T}=FK$, where in its action on electrons we may write 
\begin{equation}
S_y = \frac{1}{2 \, \mathrm{i}} \, \sum_{p} \left(c_{p,\uparrow}^\dagger \, c_{p,\downarrow}^{} - h.c.\right).
\end{equation}
Pragmatically, $F$ converts a determinant with orbital coefficients $\left(\begin{smallmatrix}\boldsymbol{C}_\uparrow \\ \boldsymbol{C}_\downarrow \end{smallmatrix}\right)$ into one with orbital coefficients $\left(\begin{smallmatrix}\boldsymbol{C}_\downarrow \\ -\boldsymbol{C}_\uparrow \end{smallmatrix}\right)$.

Together with the identity operator and complex conjugation operator, the operator product is closed:
\begin{subequations}
\begin{alignat}{2}
K \, F &= F \, K &&= \mathcal{T},
\\
K \, \mathcal{T} &= \mathcal{T} \, K &&= F,
\\
F \, \mathcal{T} &= \mathcal{T} \, F &&= -K.
\end{alignat}
\end{subequations}
While acting $K$ twice does nothing, acting $F$ or $\mathcal{T}$ on a state twice returns the same state up to a sign that is positive if the state has an even number of particles and negative if it does not.  This is because acting $F$ twice rotates each spin by $360\degree$ which is equivalent to multiplying by $-1$.

Because $K$, $F$, and $\mathcal{T}$ are not independent symmetries, we can project any two of them and the third is also automatically projected.  We take advantage of this fact below to project complex conjugation and time reversal, by projecting complex conjugation and spin flip.  We should also note that spin flip projection is equivalent to half-spin projection,\cite{smeyers1971etude,SMEYERS2000253} which eliminates half of the spin contaminants from an unrestricted wave function.  More specifically, it eliminates contaminants whose spin parity differs from that of the target.  Half-spin projection onto a singlet ($S=0$), for example, eliminates contaminants from triplet, septet, and so on (i.e. those with $S=2 \, n+1$).

\section{Complex Conjugation Projection}
\subsection{Complex Conjugation Projected Hartree-Fock}
Starting from a state $|\Psi\rangle$ which breaks complex conjugation symmetry, we can form a basis $\{|\Psi\rangle, K\,|\Psi\rangle \}$, and we can diagonalize the Hamiltonian in this basis. Since the Hamiltonian commutes with $K$, the resulting eigenvectors will naturally be eigenvectors of $K$. In this sense, we can construct the complex conjugation projection operator as
\begin{equation}
P_K = 1 + \mathrm{e}^{\mathrm{i} \, \theta} \, K.
\end{equation}
It is easy to show that
\begin{equation}
K \, P_K |\Psi\rangle = \mathrm{e}^{-\mathrm{i} \, \theta} \, P_K |\Psi\rangle,
\end{equation}
so that the projected state is an eigenfunction of $K$. 

The projection operators for spin and point group are Hermitian.  The complex conjugation projector in general is not, so we must be careful to distinguish between the complex conjugation projector acting to the right and to the left.  We define $\overleftarrow{P_K}$ to mean the projection operator acting to the left.  Given this notation, we can define the complex conjugation projected Hartree-Fock (KHF) energy as
\begin{equation}
    E[Z^{\dagger},Z] = \frac{\langle \Phi|e^{Z^\dagger} \, \overleftarrow{P_K} \, H\, P_K\, e^Z|\Phi\rangle}{\langle \Phi|e^{Z^\dagger} \, \overleftarrow{P_K} \,  P_K\, e^Z|\Phi\rangle}
\label{PHF_Ene}
\end{equation}
where the reference $|\Phi\rangle$ is chosen to be real and the complex-valued Thouless transformation\cite{Thouless1960} $Z$, given by
\begin{equation}
Z = \sum_{i,a} z_i^a \, c_a^\dagger \, c_i,
\end{equation}
is used to optimize the broken-symmetry reference. Here, $i$ and $a$ are occupied and unoccupied orbitals, respectively. We will also need to solve a $2\times 2$ CI problem to determine the coefficient $f_K = \mathrm{e}^{\mathrm{i} \, \theta}$ for each pair of $[Z^{\dagger},Z]$.

As we noted earlier, we can project both complex conjugation and time reversal (another anti-unitary symmetry) by instead projecting complex conjugation and spin-flip symmetries.  This means writing
\begin{equation}
    E[Z^{\dagger},Z] = \frac{\langle \Phi|e^{Z^\dagger} \, \overleftarrow{P_K} \, P^\dagger_F \, H\, P_F \,P_K\, e^Z|\Phi\rangle}{\langle \Phi|e^{Z^\dagger} \, \overleftarrow{P_K} \, P^\dagger_F \, P_F \, P_K\, e^Z|\Phi\rangle}.
\label{PHFEne}
\end{equation}
Here $P_F = 1 \pm F$, where the sign is $1$ for even spin (e.g. a ground-state singlet) and $-1$ for odd spin (e.g. a triplet).  We optimize this energy functional with the BFGS algorithm; the analytical gradient is included in the appendix.

Following the work of Ghassemi-Tabrizi and coworkers\cite{Shadan2020}, we could eliminate the phase in the projection operator by incorporating a phase into the optimized reference $|\tilde{\Phi}\rangle = e^Z |\Phi\rangle$.  That is, if we write
\begin{equation}
|0\rangle = \mathrm{e}^{-\mathrm{i} \, \theta/2} \, |\tilde{\Phi}\rangle,
\end{equation}
then we have
\begin{align}
P_K |\tilde{\Phi}\rangle
 &= P_K \mathrm{e}^{\mathrm{i} \, \theta/2} \, \mathrm{e}^{-\mathrm{i} \, \theta/2} |\tilde{\Phi}\rangle
\nonumber
\\
 &= P_K \mathrm{e}^{\mathrm{i} \, \theta/2} |0\rangle
\nonumber
\\
 &= \left(1 + \mathrm{e}^{\mathrm{i} \, \theta} K\right) \mathrm{e}^{\mathrm{i} \, \theta/2} |0\rangle
\nonumber
\\
 &= \mathrm{e}^{\mathrm{i} \, \theta/2} |0\rangle + \mathrm{e}^{\mathrm{i} \, \theta} \, \mathrm{e}^{-\mathrm{i} \, \theta/2} \, K |0\rangle
\nonumber
\\
 &= \mathrm{e}^{\mathrm{i} \, \theta/2} \left(1+K\right) |0\rangle = \mathrm{e}^{\mathrm{i} \, \theta/2} \, P_K |0\rangle.
\label{FixGauge} 
\end{align}
In other words, given a complex determinant $|\tilde{\Phi}\rangle$ which is projected by $P_K = 1 + \mathrm{e}^{\mathrm{i} \, \theta} K$ so that the projected state $P_K |\tilde{\Phi}\rangle$ is a $K$ eigenstate with eigenvalue $\mathrm{e}^{-\mathrm{i} \, \theta}$, we can construct an associated complex determinant $|0\rangle$ such that the projector is simply $1+K$ and the corresponding $K$ eigenvalue of $P_K |0\rangle$ is just 1.  We will take advantage of this freedom when projecting the coupled cluster equations, where it slightly simplifies the formalism.

\subsection{Connection with Other Multireference Methods}
Before we discuss complex conjugation projected coupled cluster theory, let us take a moment to consider the structure of the KHF state.  To simplify the presentation, we assume that we are projecting a complex RHF determinant that has only one (complex) occupied orbital $|\chi\rangle$.  We can write $|\chi\rangle$ in terms of two real molecular orbitals $|\phi_p\rangle$ and $|\phi_q\rangle$ as
\begin{equation}
    |\chi\rangle = \cos(\theta) \, |\phi_p\rangle + \mathrm{i} \sin(\theta) \, |\phi_q\rangle
\end{equation}
where $\theta$ is a rotation angle. The determinant is then
\begin{align}
|\chi_{\uparrow} \chi_{\downarrow} \rangle
 &= \cos^2(\theta) |\phi_{p,\uparrow} \, \phi_{p,\downarrow}\rangle
 - \sin^2(\theta) |\phi_{q,\uparrow} \, \phi_{q,\downarrow}\rangle
\\
 &+ \textrm{imaginary part}.
\nonumber
\end{align}
After projection, only the real part survives, and it takes the form of a perfect pairing generalized valence bond state,\cite{Hurley1953,Parks1958,Hay1972,Hunt1972,Goddard1973,Goddard1978} except that $|\phi_p\rangle$ and $|\phi_q\rangle$ need not be individually orthonormal (but can be).  Once we have more orbitals involved, the picture is more complicated, and in general KHF has less flexibility than a perfect pairing valence bond state for multiple electron pairs.

Complex conjugation projected Hartree-Fock is a simple non-orthogonal configuration interaction (NOCI). As such, it has the potential to describe some nearly degenerate systems.  Because it is parametrized by a single complex determinant $\mathrm{e}^Z |\Psi\rangle$, it is easy to optimize the complex orbitals defined by the Thouless parameters $Z$, avoiding any arbitrary choices one might otherwise have to make in the NOCI.  The price to pay is that complex conjugation projection, on its own, involves only two determinants.  Fortunately, it can be combined with spin and point group symmetry projection to obtain a much more complete description of the strong correlations in a molecule while still depending on only one Thouless operator $Z$.

\subsection{Complex Conjugation Projected Coupled Cluster}
After minimizing the PHF energy to get the KHF broken symmetry determinant $|\tilde{\Phi}\rangle = e^Z |\Phi\rangle$, we can use the procedure of Eqn. \ref{FixGauge} to reduce the projection operator to simply $1+K$.  We discuss the reason for this gauge-fixing step below.  We can then add coupled cluster theory to recover the dynamic correlation. We mimic the traditional coupled cluster equations to write down projected coupled cluster equations
\begin{align}
    E &= \langle 0 |H \, P_F \, P_K \, e^T| 0\rangle \label{CCEne}\\
    0 &= \langle \mu |(H-E) \, P_F \, P_K \, e^T| 0\rangle \label{CCAmp}
\end{align}
Here $\mu$ denotes excited states of the broken symmetry reference, and the cluster operator $T$ creates excitations out of $|0\rangle$. 
The key step in solving the projected coupled cluster equations is evaluating 
\begin{equation}
    \langle 0 | K \, H \, e^T |\Phi\rangle = \langle 0 | R_K \, H^*\, e^{T^*} |0\rangle
    \label{cmplx ene}
\end{equation}
where a ${}^*$ on an operator means taking the complex conjugate of its coefficients. For example, $T^\star$ has the usual form of the cluster operator but uses the complex conjugate of the cluster amplitudes.

By using the Thouless theorem and replacing $P_K$ with $\overleftarrow{P_K}$ so that we may act the projection operator to the left, we have
\begin{equation}
    \langle 0 | R_K  = \langle 0 | R_K | 0 \rangle \langle 0 | e^V
\end{equation}
where $V$ is a pure 1-body de-excitation operator. Then Eq. \ref{cmplx ene} becomes
\begin{align}
    & \langle 0 | R_K \, H^*\, e^{T^*} |0\rangle =\langle 0 | R_K | 0 \rangle \langle 0 | \, H_V^*\, e^V e^{T^*} |0\rangle, \\
    & H_V^* = e^V H^* e^{-V}.
\end{align}
One can write
\begin{equation}
\mathrm{e}^V \, \mathrm{e}^{T^\star} |0\rangle = \mathrm{e}^W |0\rangle
\end{equation}
where $W$ is a pure excitation operator.  We refer to this approach as the disentangled cluster approximation,\cite{Qiu2017} although we note that the approximation is not in writing the product of exponentials in terms of a single exponential excitation operator $W$ but in truncating $W$ (which consists of a normalization constant and all possible excitations) to a lower operator rank.  We have considered various approaches to extract an approximate $W$ from $V$ and $T$,\cite{Qiu2017,Song2022} and here we just match the power series expansion of $\mathrm{e}^W |0\rangle$ in terms of $W$ to that of $\mathrm{e}^V \mathrm{e}^{T^\star} |0\rangle$ in terms of $T^\star$.  For more details consult Ref. \citenum{Song2022}.  We note that the effects of single excitations $T_1$ can be evaluated without approximation.

Finally, we must say a few words about the projector $P_K$.  In general, the phase factor $\mathrm{e}^{\mathrm{i} \, \theta}$ for the projected Hartree-Fock and projected coupled cluster wave functions will differ.  We can solve the KCC equations while adjusting this phase or while leaving it the same as it is in KHF.  When the KCC amplitude equations are converged, we find that these two choices are equivalent.

\section{Results}
We test the proposed methods on small molecules with the cc-pVDZ basis set. Note that generally complex conjugation symmetry does not spontaneously break in Hartree-Fock, but in KHF it always does so by optimizing the mean-field state in the presence of the projection operator in what is known as the variation after projection approach.  For projected coupled cluster theory, we consider only CCSD (that is, $T$ is limited to single- and double-excitations) and we approximate the disentangled cluster operator $W$ to likewise consist only of a constant plus single- and double-excitations.  We match the two power series through second order in $W$ and $T_2$.  It would be interesting to optimize orbitals not for PHF but for PCC as well, but this requires extra effort and is beyond the scope of this work.

\begin{figure}[t]
\includegraphics[width=\columnwidth]{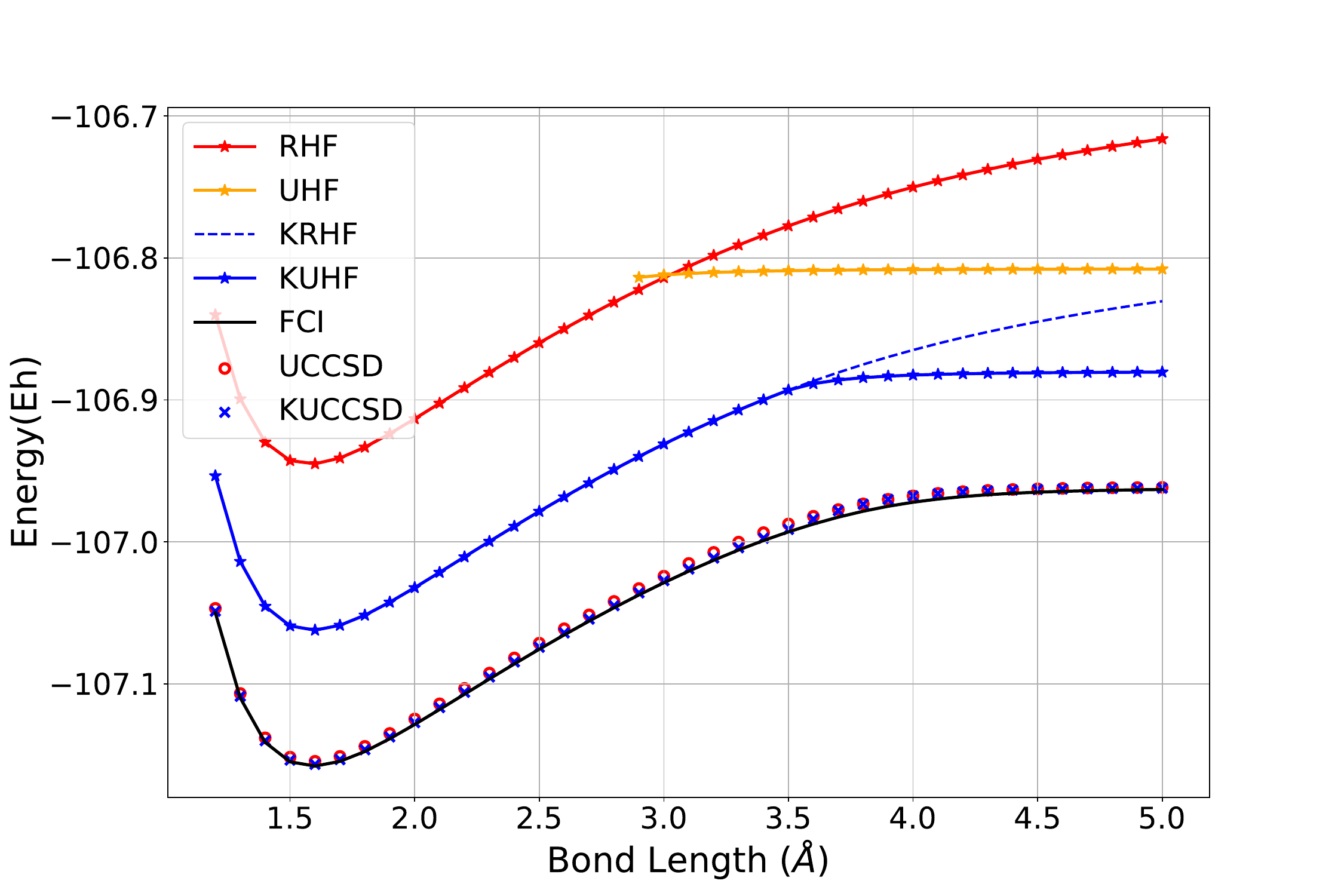} 
\caption{Energy of various HF and CCSD in LiF dissociation. 
\label{Fig:LiF}}
\end{figure}

\begin{figure}[t]
\includegraphics[width=\columnwidth]{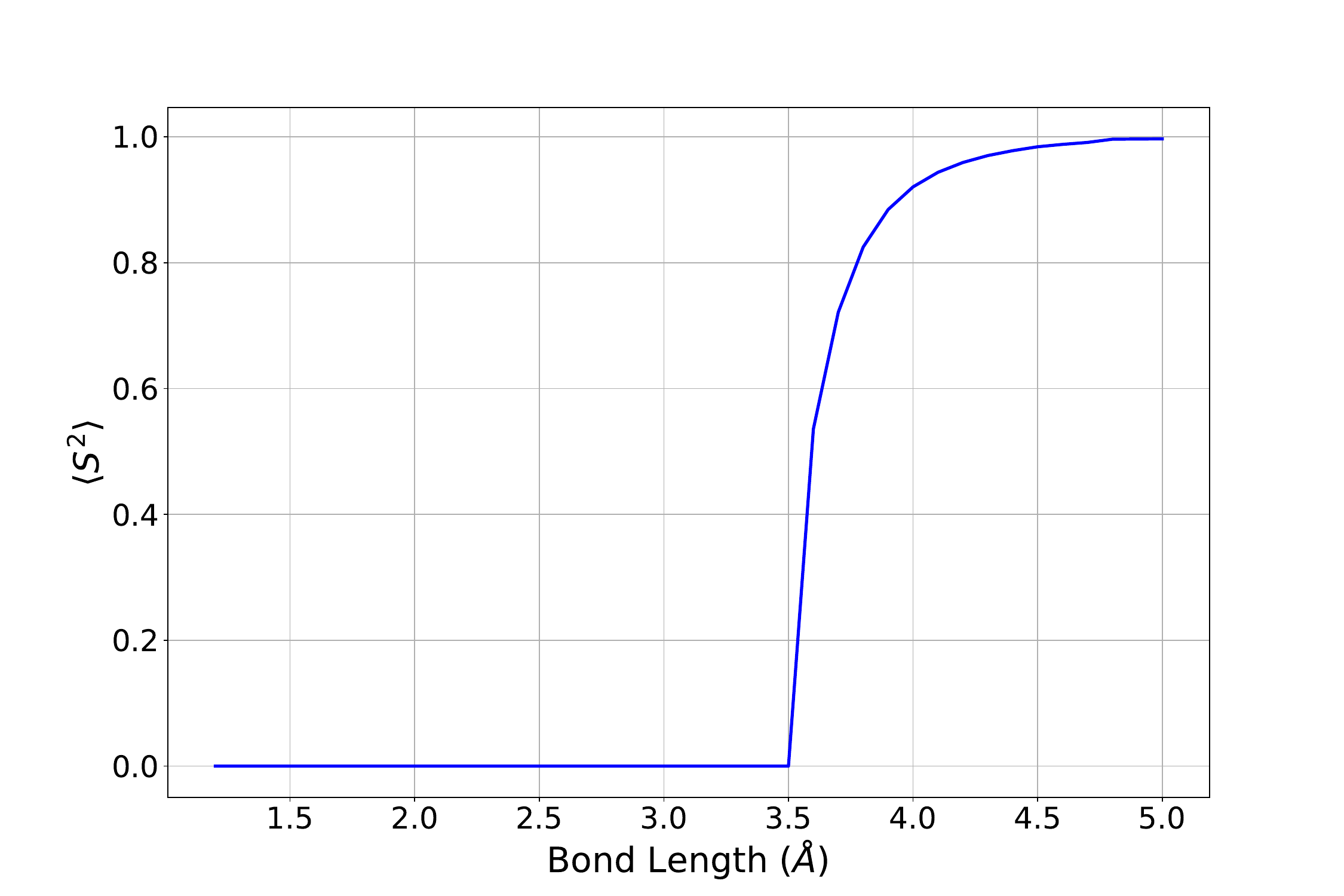} 
\caption{Expectation value of $S^2$ of the broken symmetry reference state in LiF dissociation.
\label{Fig:LiFS2}}
\end{figure}

\begin{figure}[t]
\includegraphics[width=\columnwidth]{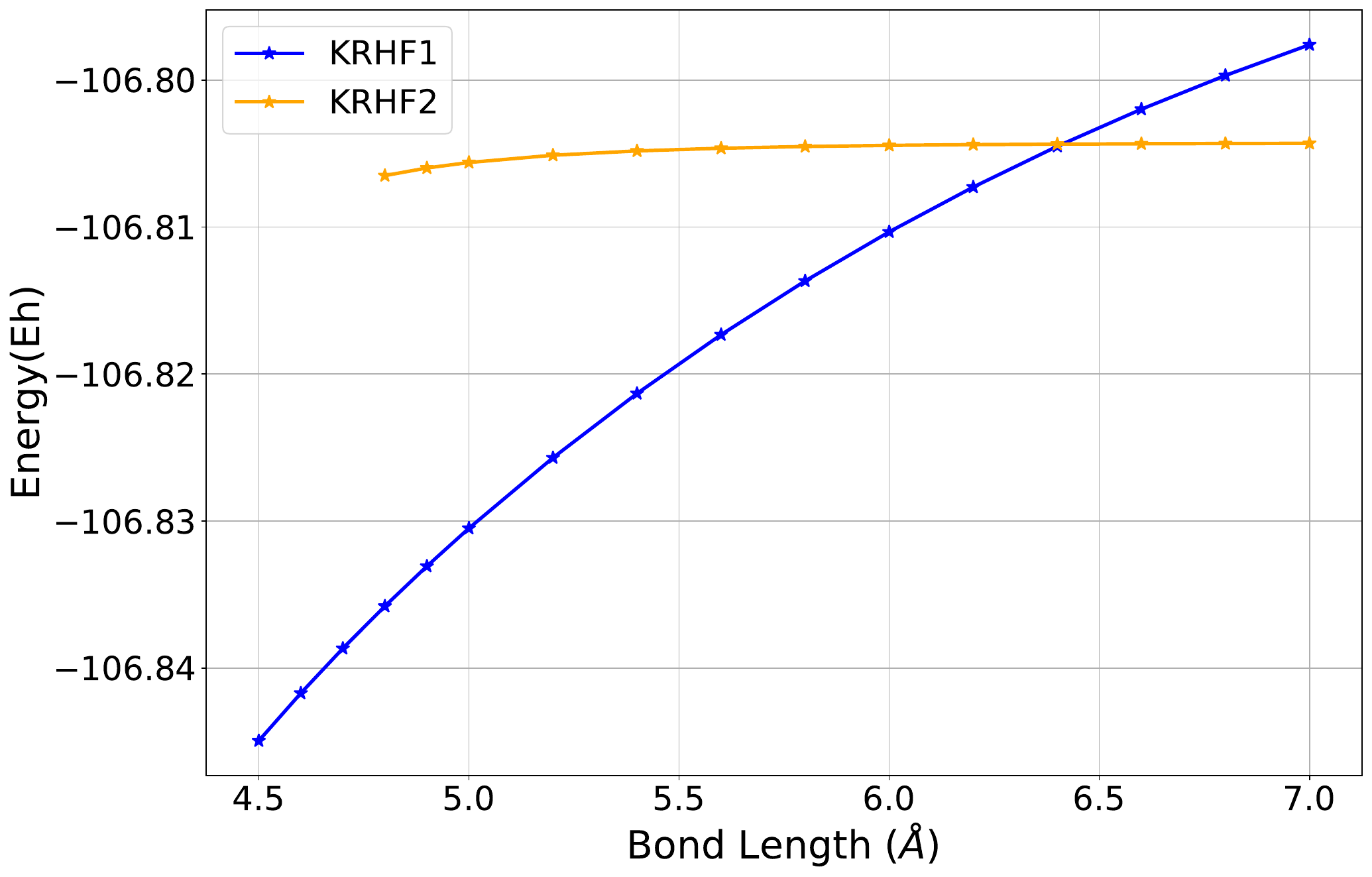}
\\
\includegraphics[width=\columnwidth]{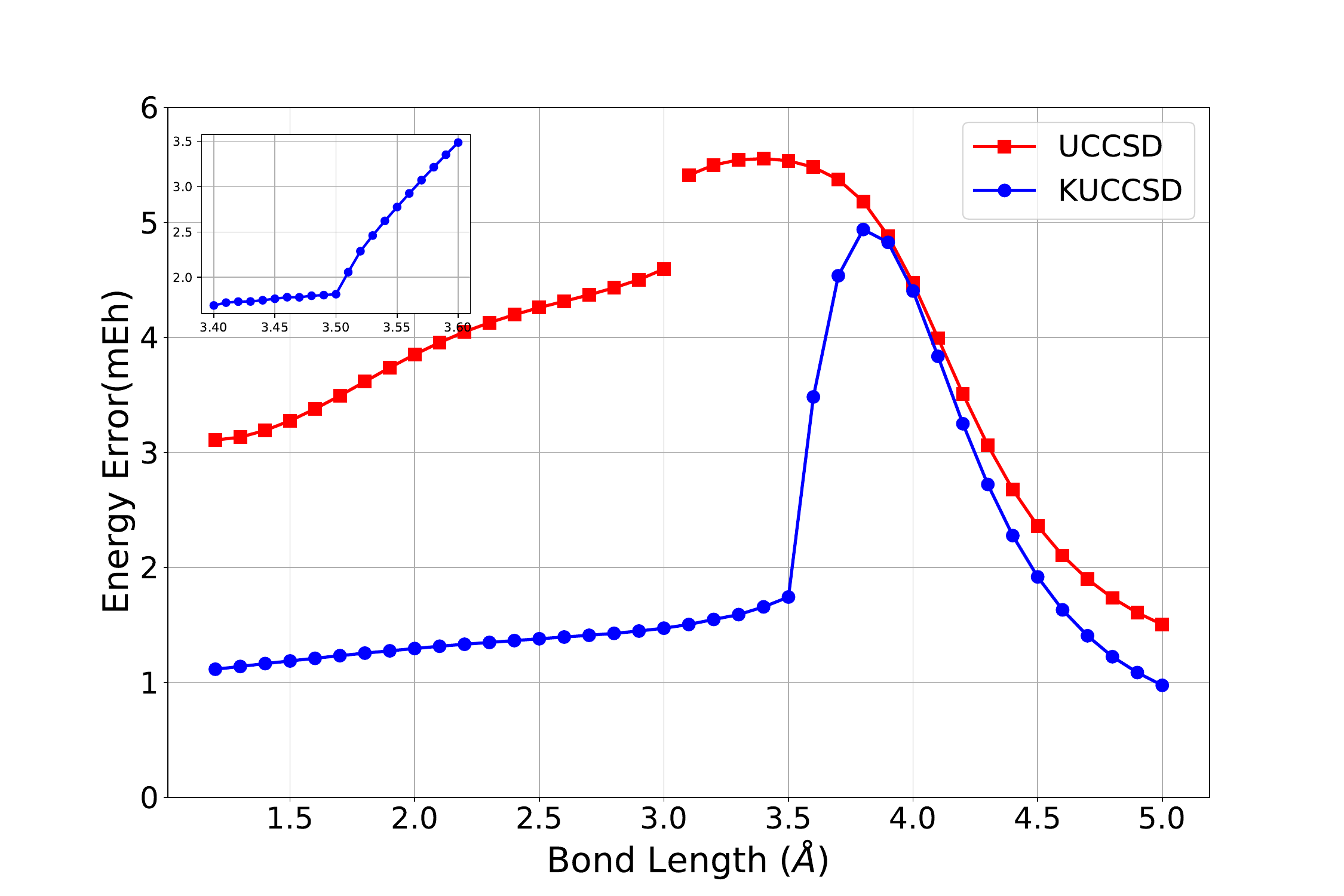}
\caption{Top panel: two KRHF solutions for the dissociation of LiF.  Bottom panel: Energy error in LiF dissociation with CCSD methods.
\label{Fig:LiFError}}
\end{figure}

\subsection{LiF Dissocation}
We start with the dissociation of LiF, which is a well-known multi-reference problem\cite{Burton2020, Carlos2019}  The molecule shows ionic character at short bond lengths, but dissociates into neutral atoms. Both configurations have $^1\Sigma_+$ symmetry, causing challenges in actual calculation. 

In this case, we allow $S^2$ symmetry to break to account for the strong correlations between the two neutral atoms and add complex conjugation projection (but not spin symmetry projection).

Figure \ref{Fig:LiF} shows the LiF dissociation curve calculated with several methods: HF and KHF, CCSD and KCCSD, and the full configuration interaction (FCI) reference.  

Let us begin with the HF and KHF data.  The ionic RHF and covalent UHF solutions cross at a bond length of around 3.0 \r{A}.  We can follow the UHF solution only slightly past the crossing point. When applying complex conjugation to UHF to get KUHF, the energy curve evolves to the correct limit at a large bond length and avoids the derivative discontinuity in the middle. Although KUHF looks much better than RHF or UHF, the Coulson–Fischer point at which spin symmetry breaks is still perceptible (see Fig. \ref{Fig:LiFS2}), although the point of spin symmetry breaking is extended from roughly 3 \r{A} to about 3.5 \r{A}.

One might wonder whether KRHF, as a two-determinant wave function, is able to correctly dissociate LiF.  This appears not to be the case (see top panel of Fig. \ref{Fig:LiFError}).  The lowest energy KRHF solution at equilibrium dissociates to ionic fragments instead of neutral ones.  There is a KRHF curve that dissociates to neutral fragments, but it is above the ionic solution except for large bond lengths and we are unable to follow it inward to less than about 4.7 \r{A}.  On the other hand, KUHF in this case appears to be size consistent (data not shown).

Now we turn to the CCSD results.  In this case, we use the RHF reference for smaller bond lengths and the UHF reference for larger bond lengths.  The fact that the reference changes explains the discontinuity observed in the CCSD curve in the bottom panel of Fig. \ref{Fig:LiFError}.  For shorter bond lengths, KCCSD is much more accurate than CCSD, but the energy error increases significantly once spin symmetry is broken, and unrestricted CCSD and KCCSD have similar errors overall.   Importantly, however, where UCCSD is not even continuous, KUCCSD yields a continuous albeit not differentiable dissociation profile (see inset in the bottom panel of Fig. \ref{Fig:LiFError}).

\begin{figure}[t]
\includegraphics[width=\columnwidth]{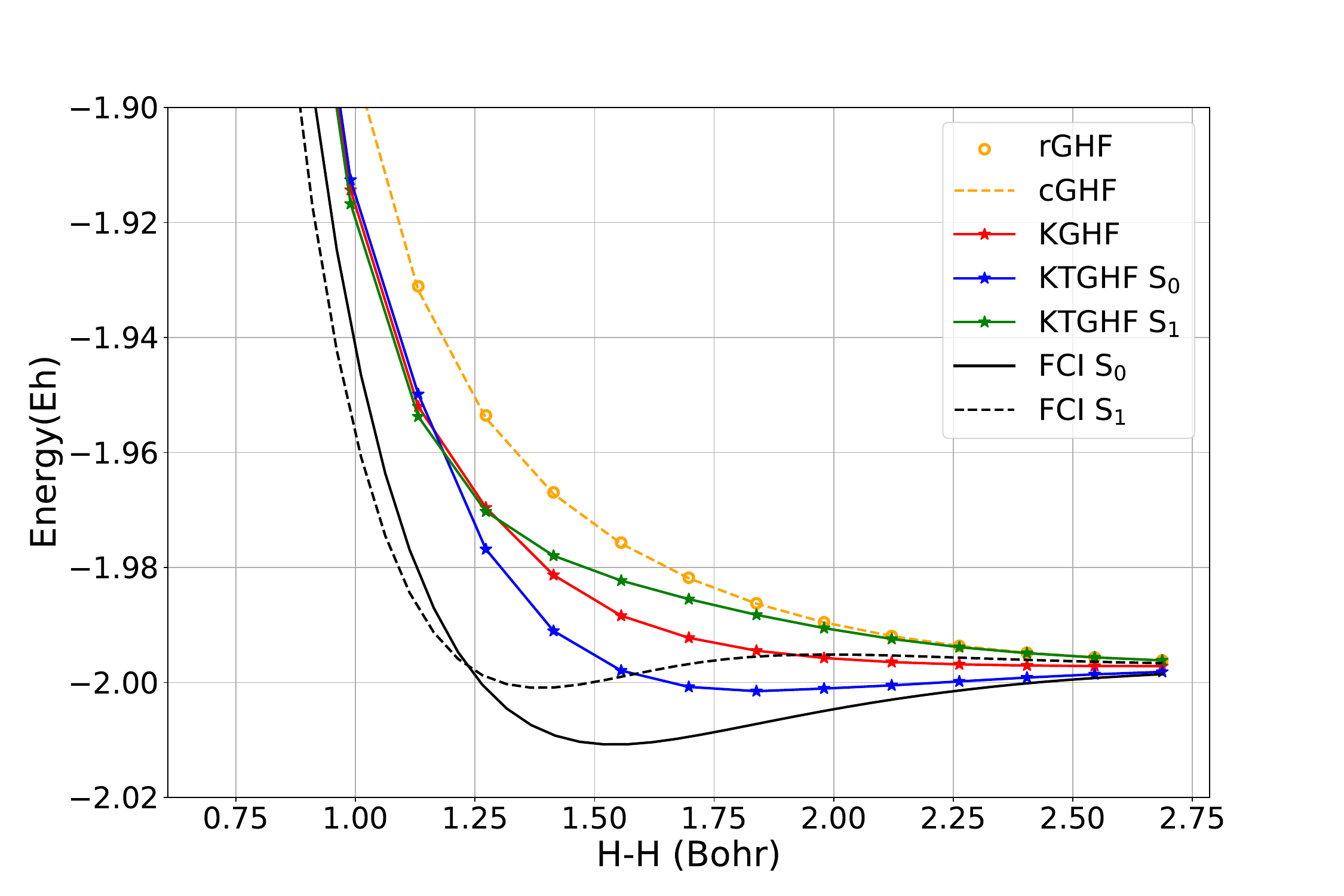} 
\\
\includegraphics[width=\columnwidth]{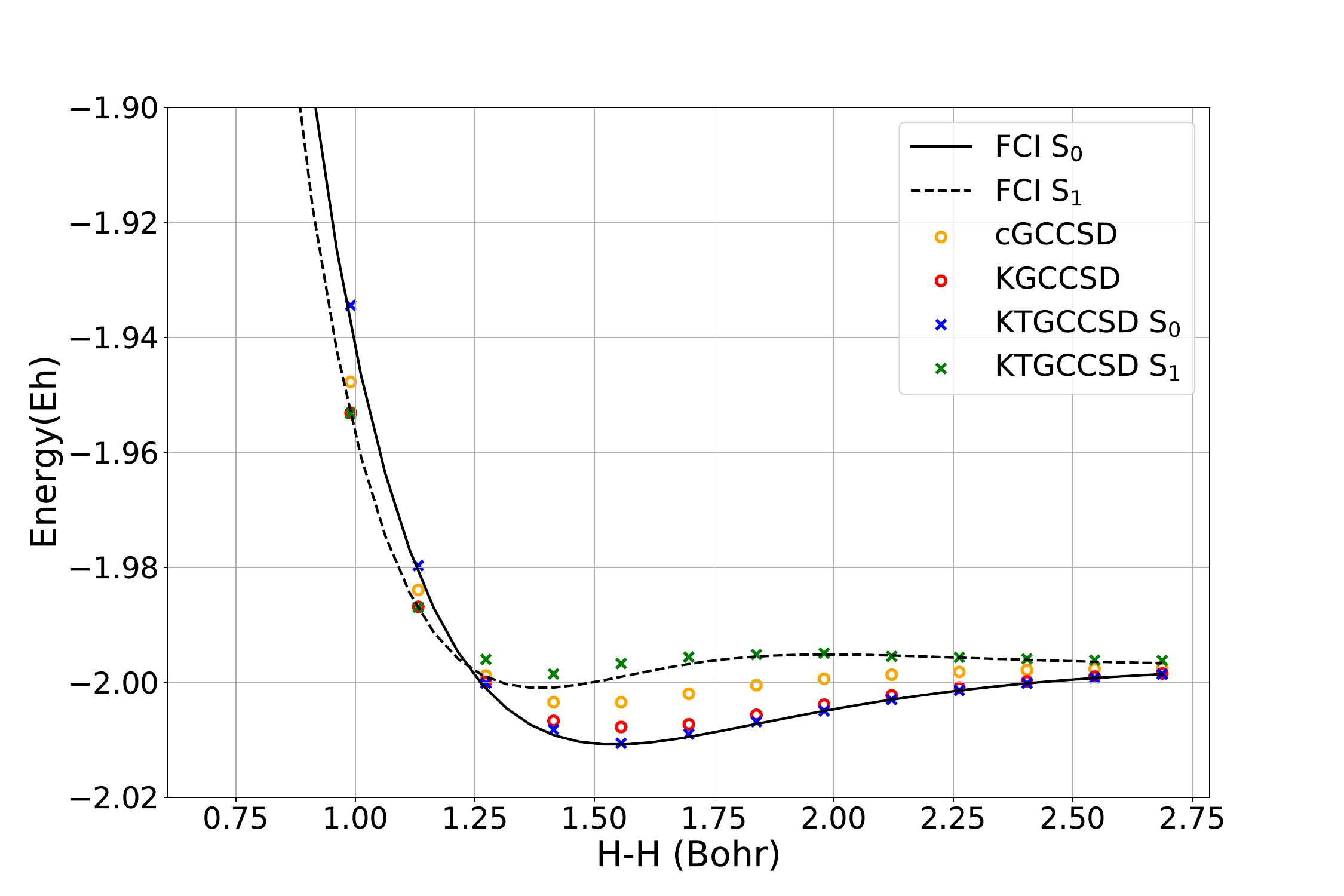}
\caption{Top panel: Energy of various HF in symmetric stretching of H$_4$ .  Bottom panel: Energy of various CC in symmetric stretching of H$_4$.
\label{Fig:H4}}
\end{figure}

\subsection{H$_4$}
Tetrahedral H$_4$ is a small molecule with spin frustration. Due to its unique geometry, one can find a real RHF, a real UHF, a real GHF, and a complex GHF solution.\cite{Tom2018}  The real RHF solution is an $S^2$ eigenstate with doubly-occupied spatial orbitals, while the real UHF has different orbitals for different spins, so respects $S_z$ symmetry but not $S^2$.  The real GHF solution breaks both $S^2$ and $S_z$ symmetries but yields coplanar magnetism, while the complex GHF solution yields a noncoplanar magnetic structure.  Except for the real RHF solution, all of the mean-field solutions we found break time-reversal symmetry, so, in this case, we restore both time reversal and complex conjugation (by, as noted, projectively restoring complex conjugation and spin-flip symmetries).  We note that paired mean-field solutions which break complex conjugation symmetry but preserve time reversal are possible in principle,\cite{Fukutome1981,Stuber2003} but we have not found them here.

In the dissociation, there is a singlet-triplet crossing at short bond length. Thus, we use $1+F$ for singlet (denoted as $S_0$ because the $S$ quantum number for a singlet is 0) and $1-F$ for triplet (denoted as $S_1$ since the $S$ quantum number for a triplet is 1) projection respectively.

\begin{figure}[t]
\includegraphics[width=\columnwidth]{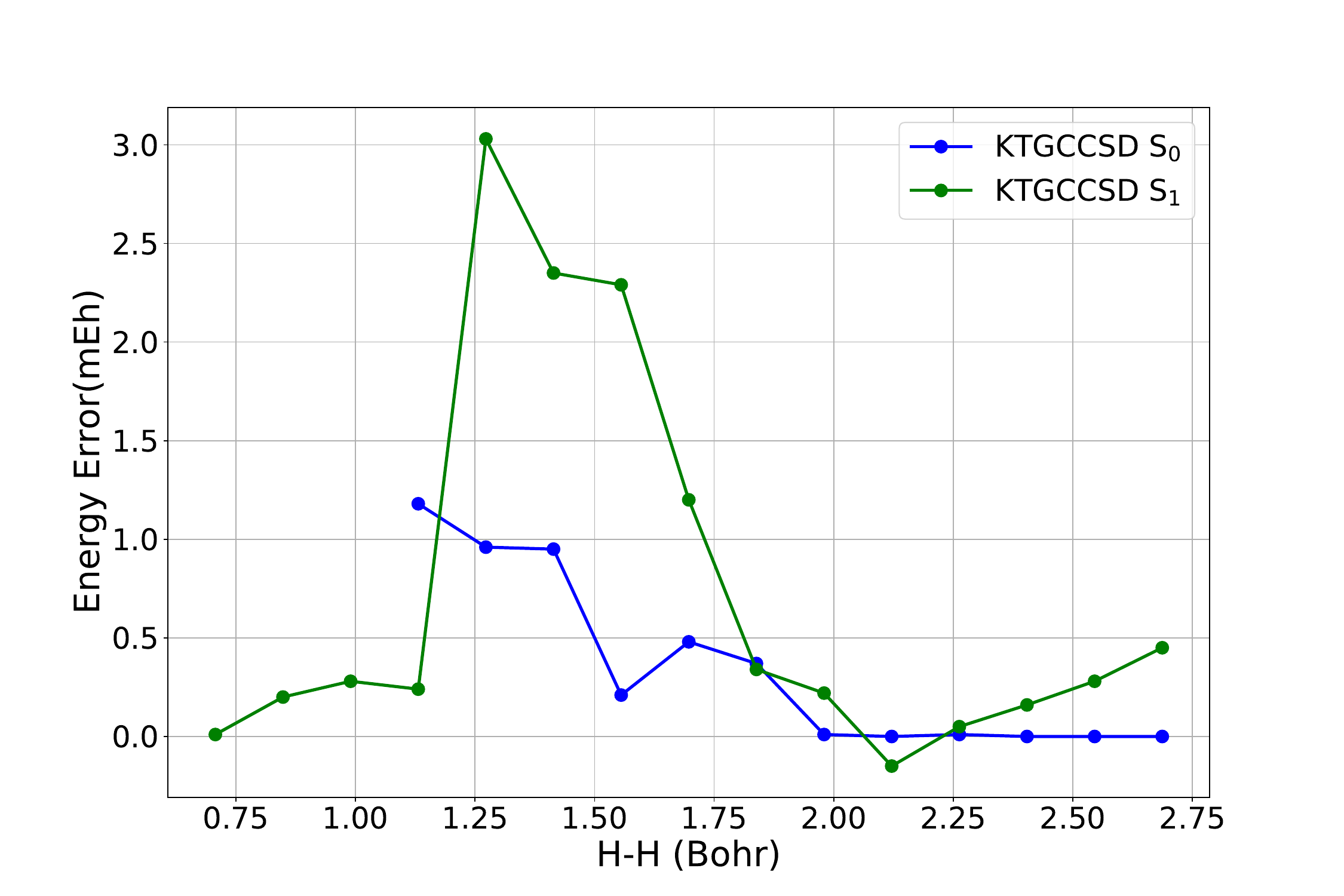} 
\caption{Energy error of projected CC in symmetric stretching of H$_4$ .
\label{Fig:H4_err}}
\end{figure}

Figure \ref{Fig:H4} shows that at the mean-field level, the system is not bonded.  Complex conjugation projected GHF (KGHF) also has no bond, though complex conjugation and singlet time-reversal projected GHF (KTGHF $S_0$)) has a very small minimum at an H-H distance of about 1.9 Bohr.  Also, we note that the crossing of singlet and triplet KTGHF (the latter denoted as KTGHF $S_1$) occurs for hydrogen-hydrogen bond lengths about 1.2 Bohr, at a slightly smaller bond length than the FCI singlet and triplet solutions cross.

As a system with only four electrons, the complex GHF-based CCSD (cGCCSD) is not too far from FCI, but the wave function switches from triplet to singlet at 1.25 Bohr, where the FCI solutions cross.  That is, the cGCCSD tracks the lower energy FCI solution rather than remaining on a single state.  Similar behavior has been seen elsewhere in the GCCSD solution for O$_2$ dissociation.\cite{JimenezHoyos2011}
Generally, KGCCSD is much closer to FCI than is GCCSD, but it also tracks the lower energy FCI solution.  Including time reversal projection not only yields improved energies but also allows us to partially control the spin state onto which we project. Figure \ref{Fig:H4_err} shows the maximum errors are about 1 and 3 mEh for singlet and triplet respectively. However, at sufficiently short bond length we are unable to fully control spin symmetry just by complex conjugation and time reversal projection since, as we noted, this amounts to only half-spin projection.  Evidently, full spin projection is needed.  We note, though, that projecting both complex conjugation and time reversal amounts to a PHF which is a linear combination of 4 determinants, while fully spin-projected GHF requires integrating over three continuous Euler angles, and would generally be expected to require a much larger number of states in the PHF.


\begin{figure}[t]
\includegraphics[width=\columnwidth]{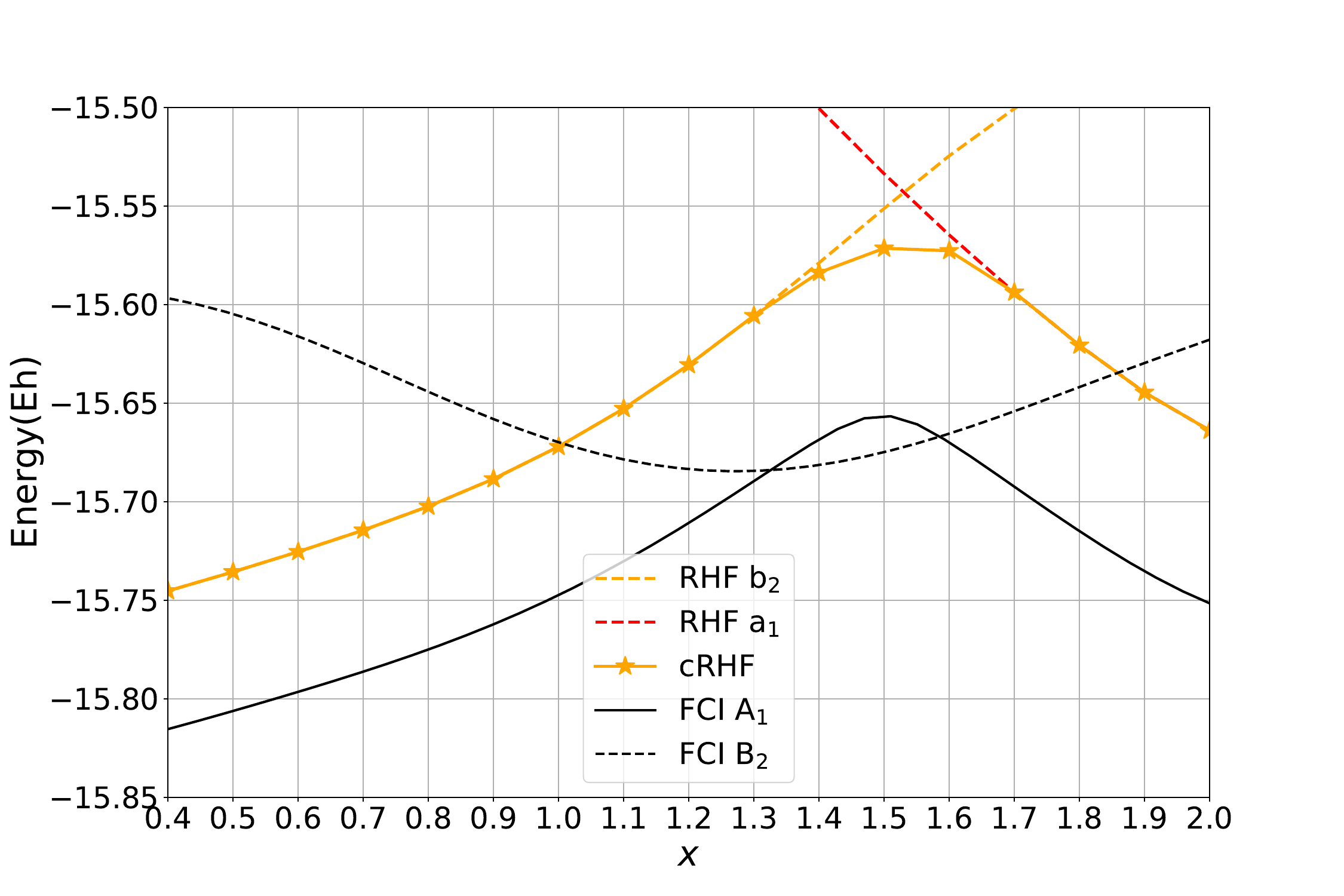} 
\\
\includegraphics[width=\columnwidth]{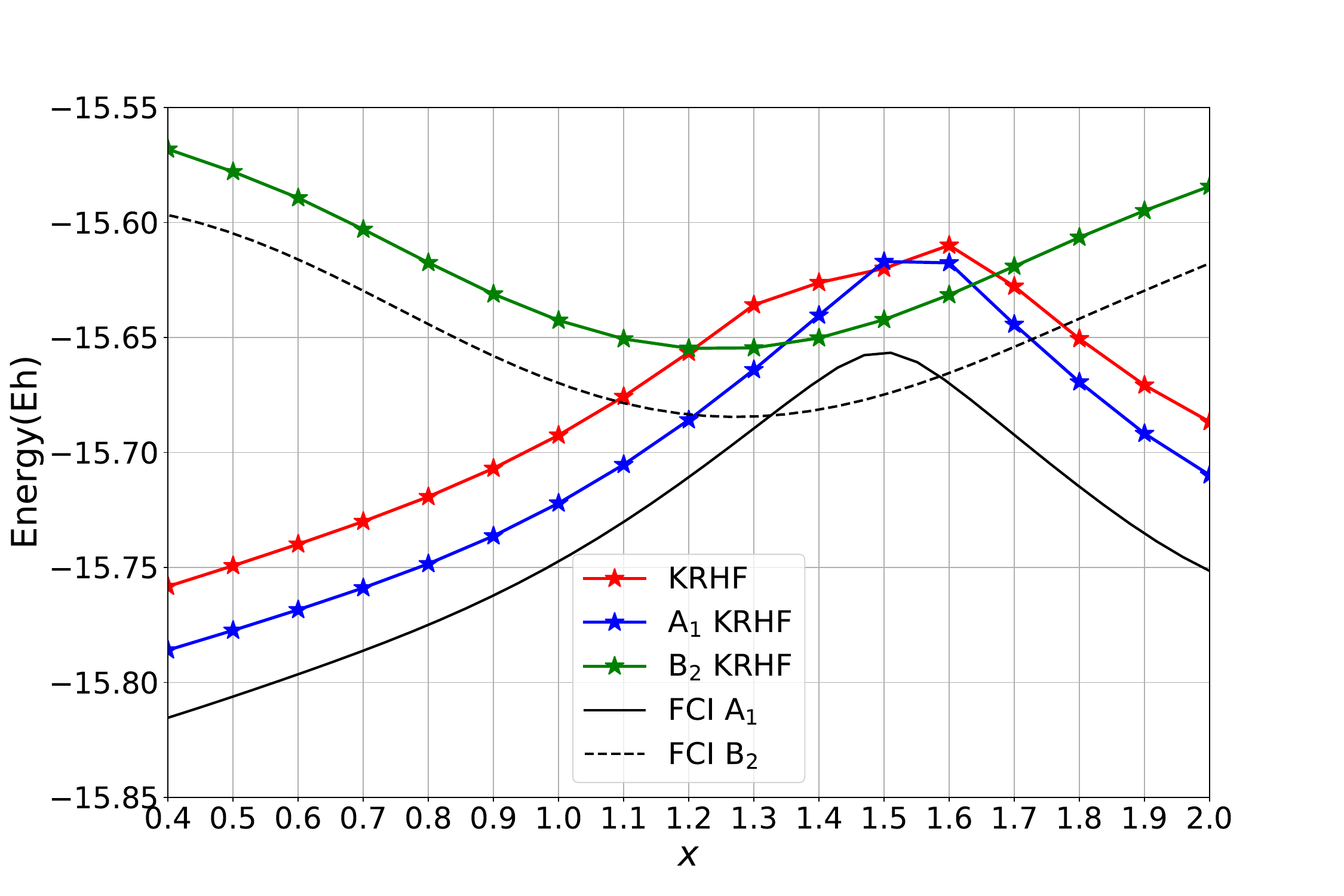} 
\\
\includegraphics[width=\columnwidth]{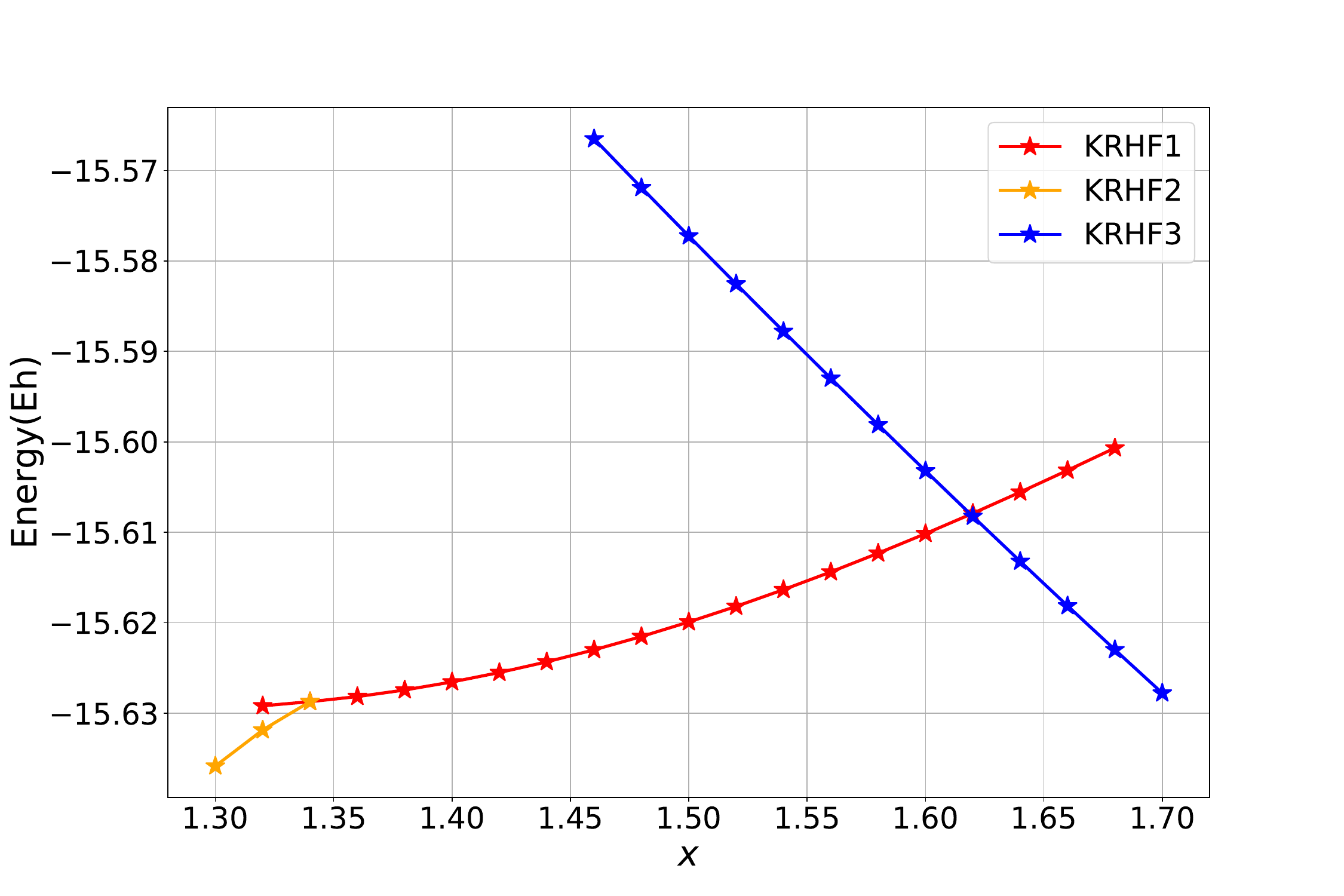} 
\caption{Total energies in the BeH$_2$ insertion.  Top panel: Comparison between HF and FCI.  Middle panel: Comparison between PHF and FCI.  Bottom panel: Energies in the three distinct KRHF solutions.
\label{Fig:BeH2}}
\end{figure}

\subsection{BeH$_2$}
Insertion of beryllium into H$_2$ is another classic example in the literature.\cite{Bartlett1983, MHG2015, Carlos2013}  We consider the C$_{2\mathrm{v}}$ geometry, where the beryllium atom is at the origin and the hydrogen atoms have coordinates $\big(x, \pm \left(1.344-0.46 \, x\right), 0\big)$ where the units are in \r{A}.

The strong correlations arise from the crossing of the frontier A$_1$ and B$_2$ molecular orbitals as $x$ evolves from 0 (the linear geometry) to 2.  At the linear geometry, the RHF ground state is of the form $(1a_1)^2 \, (2a_1)^2 \, (1b_2)^2$.  While this state has A$_1$ symmetry, we refer to it as the ``HF b$_2$'' state to denote the symmetry of the highest occupied molecular orbital.  At $x=2$, on the other hand, the RHF ground state is of the form $(1a_1)^2 \, (2a_1)^2\, (3a_1)^2$.  This state is also of A$_1$ symmetry, and we refer to it as the ``HF a$_1$'' state.  These two RHF solutions cross at $x \approx 1.5$ (see the top panel of Figure \ref{Fig:BeH2}).  Previous studies\cite{MHG2015} showed that complex RHF can smoothly connect the two configurations, but is far from the exact answer around the crossing region.  Note that complex conjugation symmetry only spontaneously breaks near $x=1.5$.

At the FCI level, meanwhile, there are three low-energy singlets.  Two are of A$_1$ symmetry and reflect the two RHF solutions we have just described, and the third is of B$_2$ symmetry and is essentially the open-shell singlet arising from occupying the frontier B$_2$ and A$_1$ orbitals once each.  Where the RHF states cross near $x=1.5$, the FCI instead has an avoided crossing, and this avoided crossing implies strong correlation.  Complicating things somewhat is that where the exact ground state is of A$_1$ symmetry across most of the insertion pathway, the B$_2$ open-shell singlet is the ground state near $x = 1.5$.

The middle panel of Figure \ref{Fig:BeH2} shows the energy of various PHF states in comparison to the FCI benchmark.  Comparing the top and middle panels reveals that, though complex conjugation projected RHF (KRHF) is far from exact, it is still notably more accurate than is RHF.  Unfortunately, KRHF has a shoulder at $x \approx 1.35$ and another at $x \approx 1.65$.  This is due to the presence of the FCI B$_2$ solution.  The complex RHF has, as its highest occupied molecular orbital, a linear combination of a$_1$ and b$_2$ character so that complex RHF has broken point group symmetry.  Complex conjugation projection tends also to restore point group symmetry, and the ground state in the region of the shoulder is of B$_2$ symmetry, not of A$_1$ symmetry as we see elsewhere.  The bottom panel of Fig. \ref{Fig:BeH2} makes this point clearly.  We see that there are three distinct KRHF solutions.  For $x \lesssim 1.35$ and $x \gtrsim 1.63$, the KRHF is of A$_1$ symmetry corresponding to the two distinct A$_1$ RHF solutions.  For $x$ between these two values, KRHF instead gives a B$_2$ state.  We can follow these three distinct KRHF solutions beyond the point at which they are the KRHF ground state, but at some point we are unable to locate them.

The solution to this problem is to include point group symmetry projection along with complex conjugation projection.  These give us the curves marked in the middle panel of Fig. \ref{Fig:BeH2} as A$_1$ KRHF and B$_2$ KRHF, where we have projected complex conjugation and point group symmetry together and optimized the reference determinant accordingly.  Including point group projection yields well-behaved curves which are nearly parallel to the correspinding FCI references.

\begin{figure}[t]
\includegraphics[width=\columnwidth]{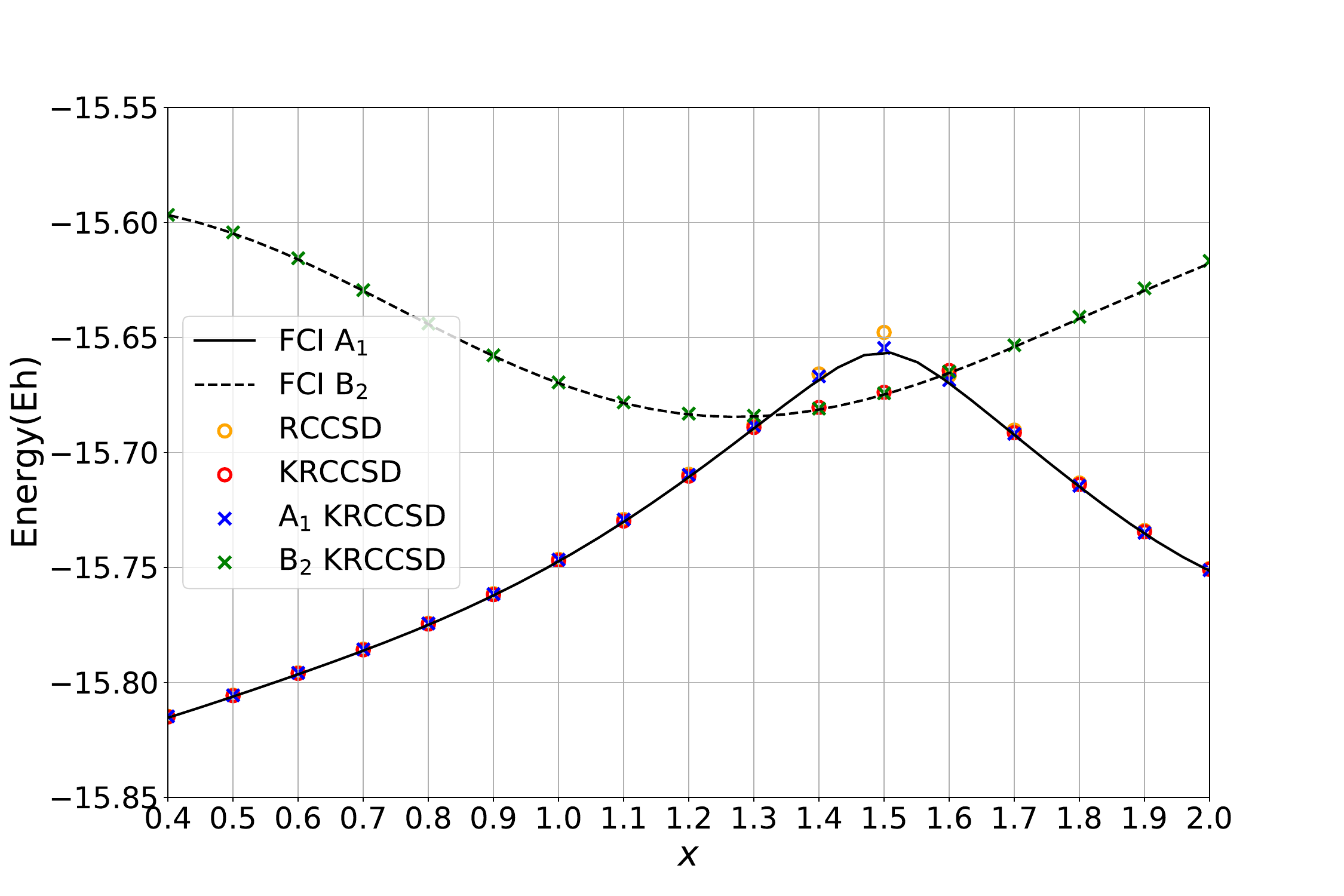} 
\\
\includegraphics[width=\columnwidth]{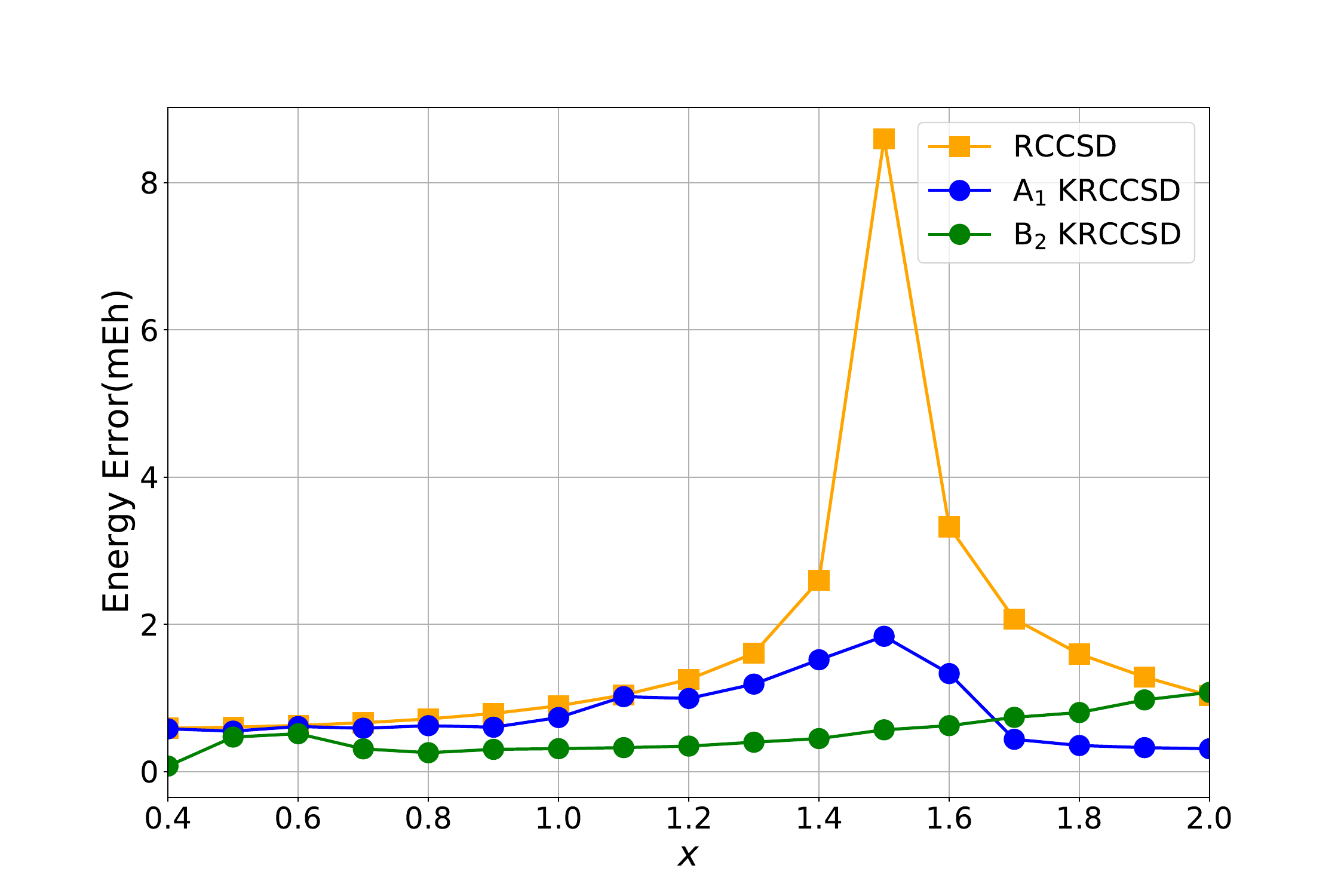} 
\caption{Coupled cluster energies in the BeH$_2$ insertion.  Top panel: Total energies, compared to FCI.  Bottom panel: Energy error of CCSD and projected CCSD compared to FCI.
\label{Fig:BeH2_2}}
\end{figure}

Figure \ref{Fig:BeH2_2} shows the total energies (top panel) and errors with respect to FCI (bottom panel) for various coupled cluster methods.  The restricted CCSD (RCCSD) data are obtained by doing (real) RCCSD based on the two A$_1$ RHF determinants and taking the lower of the two energies at each point.  Since this RCCSD has A$_1$ symmetry everywhere, we compare it to the A$_1$ FCI, and we see that it is fairly accurate except in the immediate vicinity of $x=1.5$ where the correlations are strongest.  The complex-conjugation projected RCCSD (KRCCSD) curve is likewise obtained by doing KRCCSD curves starting from the distinct KRHF solutions and taking the lowest.  Like RCCSD, it is highly accurate, but because the KRHF reference is of B$_2$ symmetry for intermediate values of $x$, the KRCCSD curve approximates the FCI B$_2$ state in this region.  Including point group and complex conjugation projection to get A$_1$ KRCCSD or B$_2$ KRCCSD gives results which agree with the corresponding FCI with errors less than 2 mEh.

\section{Discussion and Conclusion}
\subsection{Discussion}
This work shows a systematic way to restore complex conjugation symmetry at both the Hartree-Fock and coupled cluster levels.  It is easy to check that when the symmetry is not broken, these theories go back to traditional HF and CC. The disentangled cluster formalism and power series approximation for the discrete symmetry yield satisfactory results even though the symmetries involved are discrete instead of continuous.

On the other hand, as discussed elsewhere,\cite{Qiu2018,Tsuchimochi2018,Tsuchimochi2019, Song2022} projected CC has a problem with what we call gauge invariances.  These are numerical invariances of the algebraic equations that leave observables unchanged. Although gauge invariances do not affect the final result, they still make the PCCSD result not unique when the projection is approximated (by, for example, truncating the operator $W$ to exclude triple excitations and higher).  Partial solutions have been proposed,\cite{Qiu2018,Tsuchimochi2018,Tsuchimochi2019} but no fully satisfactory means of eliminating them has been found.

In this work, we choose the projection operator to be $1+K$ to fix the gauge in complex conjugation, but whether it is the best choice still needs to be determined. When introducing spin-flip symmetry, we have an extra spin gauge mode, and this mode causes trouble in the convergence of the PCC equations. The spin-flip projection only contains two operators, but we need to use large damping factors to slow down changes in CC amplitudes which otherwise oscillate dramatically.

\subsection{Conclusions}
Complex wave functions are useful tools in describing degenerate systems and contain richer information than do real wave functions. With proper projection, we can eliminate the unphysical part of complex wave functions and achieve superior results. Coupled cluster is a sound wave function that captures dynamic correlation, so combining complex conjugation projection and coupled cluster is natural. Benefiting from the projected coupled cluster theory, we can build an effective method to implement complex conjugation projected CCSD. Also, we add spin flip projection under the same framework, which enables us to restore time-reversal symmetry as well. These projected wave functions successfully solve some well-known multi-reference problems and achieve results of roughly chemical accuracy without requiring a description of higher particle-hole excitations beyond doubles.

\begin{acknowledgments}
This work was supported by the U.S. National Science Foundation under Grant No. CHE-1762320. G.E.S. is a Welch Foundation Chair (Grant No. C-0036).
\end{acknowledgments}

\appendix
\section{Gradient for Projected HF}
In this appendix, we detail the optimization of symmetry-projected HF states based on a Thouless parametrization. We treat the Thouless parameters $z$ and $z^\star$ in Eqn \ref{PHFEne} as independent variables, and the gradient becomes
\begin{equation}
    G_{ai} = \frac{\partial E}{\partial z^*_{ai}} = \frac{\langle \Phi|c_i^\dagger c_a \, e^{Z^\dagger} \, \overleftarrow{P_K} \, P^\dagger_F \, (H-E)\, P_F \,P_K\, e^Z|\Phi\rangle}{\langle \Phi|e^{Z^\dagger} \, \overleftarrow{P_K} \, P^\dagger_F \, P_F \, P_K\, e^Z|\Phi\rangle}.
\end{equation}
Since we choose the reference state to be real, the complex conjugation operator will only change the parameters $z, z^*$ and leave the reference unchanged. 

The energy gradient is made of two parts. The kernel for the overlap-like part is 
\begin{equation}
     \frac{\langle \Phi|c_i^\dagger c_a \, e^{Z^\dagger} \, R \,e^Z|\Phi\rangle}{\langle \Phi|e^{Z^\dagger} \, R \, e^Z  |\Phi\rangle} = \rho_{ai}
\end{equation}
and that for the Hamiltonian-like part is
\begin{equation}
    \frac{\langle \Phi|c_i^\dagger c_a \, e^{Z^\dagger} \, H \, R \,e^Z|\Phi\rangle}{\langle \Phi| e^{Z^\dagger} \, R \, e^Z  |\Phi\rangle} = h \, \rho_{ai} + [(1-\rho) \, (h + \Gamma) \, \rho]_{ai}.
\end{equation}
Here $R$ is the operator in the projection operators, and $h$ and $\Gamma$ are defined as
\begin{align}
    & h = \frac{\langle \Phi| e^{Z^\dagger} \, H \, R \,e^Z|\Phi\rangle}{\langle \Phi| e^{Z^\dagger} \, R \, e^Z  |\Phi\rangle}, \\
    & \Gamma_{ik} = \sum_{jl} V_{ijkl} \, \rho_{lj}.
\end{align}

\end{document}